\documentclass[twocolumn, onecolappendix]{aastex62}

\usepackage{datetime}
\newdateformat{myformat}{\monthname[\THEMONTH] \THEYEAR}

\usepackage{xcolor}
\usepackage{graphicx}
\usepackage{subfigure}
\usepackage{tikz}

\usepackage{listings}

\date{\myformat\formatdate{-1}{11}{2018}}

\usepackage{amsmath, amssymb, amsthm}
\usepackage{bm}

\newcommand{\vect}[1]{\mathbf{#1}}
\newcommand{\tens}[1]{\mathbf{#1}}
\renewcommand{\div}{\bm{\nabla \cdot}}
\newcommand{\grad}{\bm{\nabla}}

\newcommand{\refereecorrection}[1]{\textcolor{black}{#1}}
\newcommand{\refereecorrectionTwo}[1]{\textcolor{black}{#1}}

\begin{document}
\title{A high performance and portable all-Mach regime flow solver
  code with well-balanced gravity. Application to compressible
  convection.}
\author{Thomas Padioleau}
\affiliation{Maison de la Simulation, CEA, CNRS, Univ. Paris-Sud,
  UVSQ, Universit\'{e} Paris-Saclay, 91191 Gif-sur-Yvette, France}
\email{thomas.padioleau@cea.fr}
\author{Pascal Tremblin}
\affiliation{Maison de la Simulation, CEA, CNRS, Univ. Paris-Sud,
  UVSQ, Universit\'{e} Paris-Saclay, 91191 Gif-sur-Yvette, France}
\author{Edouard Audit}
\affiliation{Maison de la Simulation, CEA, CNRS, Univ. Paris-Sud,
  UVSQ, Universit\'{e} Paris-Saclay, 91191 Gif-sur-Yvette, France}
\author{Pierre Kestener}
\affiliation{Maison de la Simulation, CEA, CNRS, Univ. Paris-Sud,
  UVSQ, Universit\'{e} Paris-Saclay, 91191 Gif-sur-Yvette, France}
\author{Samuel Kokh}
\affiliation{DEN/DANS/DM2S/STMF, CEA Saclay, 91191 Gif-sur-Yvette,
  France}
\begin{abstract}
  Convection is an important physical process in astrophysics
  well-studied using numerical simulations under the Boussinesq and/or
  anelastic approximations. However these approaches reach their
  limits when compressible effects are important in the high Mach flow
  regime, e.g.\ in stellar atmospheres or in the presence of accretion
  shocks.\par{}
  In order to tackle these issues, we propose a new high performance
  and portable code, called ``ARK'' with a numerical solver
  well-suited for the stratified compressible Navier-Stokes equations.
  We take a finite volume approach with machine precision conservation
  of mass, transverse momentum and total energy. Based on previous
  works in applied mathematics we propose the use of a low Mach
  correction to achieve a good precision in both low and high Mach
  regimes. The gravity source term is discretized using a
  well-balanced scheme in order to reach machine precision hydrostatic
  balance. This new solver is implemented using the Kokkos library in
  order to achieve high performance computing and portability across
  different architectures (e.g.\ multi-core, many-core, and
  GP-GPU).\par{}
  We show that the low-Mach correction allows to reach the low-Mach
  regime with a much better accuracy than a standard Godunov-type
  approach. The combined well-balanced property and the low-Mach
  correction allowed us to trigger Rayleigh-B\'{e}nard convective
  modes close to the critical Rayleigh number. Furthermore we present
  3D turbulent Rayleigh-B\'{e}nard convection with low diffusion using
  the low-Mach correction leading to a higher kinetic energy power
  spectrum. These results are very promising for future studies of
  high Mach and highly stratified convective problems in astrophysics.
\end{abstract}

\section*{Introduction}
The study of convection is an active topic of research in the
astrophysics community because of its major role in different
mecanisms such as heat transport in solar and stellar
interiors~\citep{spruit_solar_1990}, mixing of
elements~\citep{pinsonneault_mixing_1997} and
dynamo~\citep{charbonneau_solar_2014}. As these mecanisms play a role
in the estimation of the lifetime of these objects it is of great
importance for stellar evolution theory.\par{}
Different approximations have been developed to ease the study of
convection. The Boussinesq and the anelastic approximations simplify
the Navier-Stokes system by getting rid of acoustic waves and keeping
buoyancy effects. In practice these approximations are derived by
looking at the equations satisfied by small perturbations near a
reference state~\citep{spiegel_boussinesq_1960}. The Boussinesq
approximation is quite restrictive as it is valid for a small layer of
the reference state, such that the flow can be considered
incompressible. On the other hand the anelastic approach allows to
have a larger scale height by keeping the density stratification of
the reference state~\citep{gilman_compressible_1981}. Another way to
understand these approximations is to consider the flow regime in
terms of the Mach number $\mathrm{Ma}$. As it is shown
in~\cite{mentrelli_modelling_2018}, these approximations can be
recovered by considering low-Mach asymptotic limits of the
Navier-Stokes system. The Froude number, defined as the
non-dimensional ratio of kinetic energy to gravitational energy,
characterizes the influence of gravity in the flow. By taking into
account different Froude regimes, they recover the incompressible, the
Boussinesq and the anelastic models.
From a numerical point of view the removal of the acoustics waves in
these models is quite attractive because it allows to have larger time
steps. The anelastic model has been successfully implemented in
different codes like Rayleigh~\citep{featherstone_spectral_2016} or
Magic~\citep{gastine_effects_2012} and it is widely used in the
community~\citep[see][]{glatzmaier_introduction_2017}. \refereecorrection{We
  can also mention the \uppercase{Maestro}
  code~\citep[see][]{nonaka_maestro:_2010} which uses an extended
  version of the anelastic model. The velocity constraint takes into
  account the time variation of pressure. However these approaches
  present some drawbacks.} The addition of new physics and source
terms to the model is difficult, one has to derive another asymptotic
model to take the new physics into account in the anelastic
regime~\citep[see][]{mentrelli_modelling_2018}. Furthermore one has to
be careful that the simulation stays in the regime of validity of the
model (especially in the Boussinesq regime). Finally a numerical
difficulty is the parallelization of those codes. They usually use
pseudo-spectral methods for which it is more difficult to achieve a
good scalability~\citep[e.g.\ need to use pencil-type domain
decomposition][]{featherstone_spectral_2016}.\par{}
We chose to take a more flexible approach by solving the full
compressible Navier-Stokes system, as in the MUSIC
code~\citep{viallet_towards_2011, goffrey_benchmarking_2017} but with
a collocated finite volume solver \refereecorrection{instead of using
  a staggered grid}. Different discretization techniques of the Euler
system are used in the astrophysics community. We can classify them in
various ways. One way is to separate SPH techniques from grid-based
techniques. Furthermore grid-based approaches can be divided in
different families, finite difference, finite element and finite
volume. The finite volume method is of particular interest because of
its natural property of being conservative and to capture shocks and
discontinuities. Designing a finite volume scheme essentially resides
in the definition of a numerical flux, numerical counterpart of the
physical flux. A widely used family of fluxes is the
Godunov~\citep[see][]{godunov_finite_1959} flux which is the flux of
the --- usually approximate --- Riemann problem between two neighbour
cells.\par{}
However we have to face multiple numerical difficulties with this
approach. Compressible solvers and mainly Godunov-type solvers are
known to have an excessive amount of numerical diffusion in the
low-Mach regime which make them unusable in this
regime~\citep[see][]{guillard_behaviour_1999,
  dellacherie_analysis_2010, miczek_new_2015, chalons_all-regime_2016,
  barsukow_numerical_2017}. In this regime, in which flows are
smoother, considering Riemann problems at interfaces is not
adapted. Indeed in the work of~\cite{miczek_new_2015} they show that
part of the kinetic energy is dissipated into internal energy whereas
it should be conserved. \refereecorrection{To tackle this issue they
  propose a preconditionned Roe scheme to remove the numerical
  diffusion. Secondly, hydrodynamics and gravity are usually
  discretized independently from each other.} In the case of highly stratified medium,
the numerical scheme does not maintain the hydrostatic equilibrium and
produces spurious flows that pollutes the simulation. Different
approaches have been investigated to solve this issue both for the
Euler and the shallow water equations. In~\cite{leroux_schema_1994},
they rewrite the Euler system as a fully conservative system by
defining an hydrostatic pressure satisfying a conservation
law. In~\cite{chandrashekar_second_2015} they use a variable
reconstruction by taking advantage of the equilibrium
profile. In~\cite{chalons_godunov-type_2010, vides_godunov-type_2014,
  chalons_large_2017} they incorporate the source term in the Riemann
problem itself allowing to compensate pressure gradients at the
interface. As in~\cite{leroux_schema_1994}, authors
from~\cite{chertock_well-balanced_2018} also propose to discretize the
Euler system with gravity as a fully conservative system but using
global fluxes and a reconstruction on equilibrium variables. Finally
the last numerical difficulty is the time step in the low-Mach
regime. Because of the stability condition involving the fast acoustic
waves, the time step becomes very small compared to the material
transport timescale. It can either be resolved using a full implicit
approach as in the MUSIC code~\cite{viallet_towards_2011,
  goffrey_benchmarking_2017}, or by using an implicit-explicit (IM-EX)
approach in which only the system with fast acoustic waves is solved
implicitly~\citep{chalons_all-regime_2016, chalons_large_2017}.\par{}
Following the original work of~\cite{chalons_all-regime_2016}
and~\cite{chalons_large_2017} we use an acoustic-transport
splitting. In~\cite{chalons_all-regime_2016} they derive a finite
volume scheme of the Euler system on unstructured mesh. This scheme
uses an acoustic splitting to separate acoustic waves from material
ones. In the low Mach regime, this translates to a splitting between
fast waves and slow waves. In the low Mach regime, the fast waves can
be treated with an implicit solver to get rid of the restrictive
stability condition. Then in the work of~\cite{chalons_large_2017},
the scheme has been adapted to shallow water equations with a source
term which is the topography. This source term is added in the
equivalent acoustic subsystem to obtain a well-balanced
scheme. \refereecorrection{In this paper we adapt their approach for
  the Euler system by taking care of the discretization of the energy
  equation.}\par{}
The paper is organized as follows. In
Section~\ref{sec:navier_stokes_equations} we briefly recall the
compressible model we use to study convection, i.e.\ the Navier-Stokes
equations with gravity. In Section~\ref{sec:numerical_scheme} we
present the derivation of the well-balanced and all-regime numerical
scheme using a splitting approach between an acoustic step and a
transport step both solved explicitly in this work. In
Section~\ref{sec:implementation} we present some implementation
features about the
``ARK''~\refereecorrection{\footnote{\url{https://gitlab.erc-atmo.eu/erc-atmo/ark},
    version v1.0.0}} code in particular the Kokkos library used for
the shared memory parallelization. We also give some performance
results. Finally in Section~\ref{sec:numerical_results} we present
different numerical test cases illustrating the importance of the
low-Mach correction and the well-balanced discretization of gravity.
\section{Navier-Stokes equations}\label{sec:navier_stokes_equations}
\begin{figure*}
  \centering
\ifx\du\undefined
  \newlength{\du}
\fi
\setlength{\du}{15\unitlength}
\begin{tikzpicture}[even odd rule]
\pgftransformxscale{1.000000}
\pgftransformyscale{-1.000000}
\definecolor{dialinecolor}{rgb}{0.000000, 0.000000, 0.000000}
\pgfsetstrokecolor{dialinecolor}
\pgfsetstrokeopacity{1.000000}
\definecolor{diafillcolor}{rgb}{1.000000, 1.000000, 1.000000}
\pgfsetfillcolor{diafillcolor}
\pgfsetfillopacity{1.000000}
\pgfsetlinewidth{0.100000\du}
\pgfsetdash{}{0pt}
\pgfsetmiterjoin
{\pgfsetcornersarced{\pgfpoint{0.000000\du}{0.000000\du}}\definecolor{diafillcolor}{rgb}{1.000000, 1.000000, 1.000000}
\pgfsetfillcolor{diafillcolor}
\pgfsetfillopacity{1.000000}
\fill (32.058800\du,23.140000\du)--(32.058800\du,25.040000\du)--(39.741300\du,25.040000\du)--(39.741300\du,23.140000\du)--cycle;
}{\pgfsetcornersarced{\pgfpoint{0.000000\du}{0.000000\du}}\definecolor{dialinecolor}{rgb}{0.000000, 0.000000, 0.000000}
\pgfsetstrokecolor{dialinecolor}
\pgfsetstrokeopacity{1.000000}
\draw (32.058800\du,23.140000\du)--(32.058800\du,25.040000\du)--(39.741300\du,25.040000\du)--(39.741300\du,23.140000\du)--cycle;
}
\definecolor{dialinecolor}{rgb}{0.000000, 0.000000, 0.000000}
\pgfsetstrokecolor{dialinecolor}
\pgfsetstrokeopacity{1.000000}
\definecolor{diafillcolor}{rgb}{0.000000, 0.000000, 0.000000}
\pgfsetfillcolor{diafillcolor}
\pgfsetfillopacity{1.000000}
\node[anchor=base,inner sep=0pt, outer sep=0pt,color=dialinecolor] at (35.900050\du,24.285000\du){Gravitational energy};
\pgfsetlinewidth{0.100000\du}
\pgfsetdash{}{0pt}
\pgfsetmiterjoin
{\pgfsetcornersarced{\pgfpoint{0.000000\du}{0.000000\du}}\definecolor{diafillcolor}{rgb}{1.000000, 1.000000, 1.000000}
\pgfsetfillcolor{diafillcolor}
\pgfsetfillopacity{1.000000}
\fill (52.325000\du,23.140000\du)--(52.325000\du,25.040000\du)--(58.330000\du,25.040000\du)--(58.330000\du,23.140000\du)--cycle;
}{\pgfsetcornersarced{\pgfpoint{0.000000\du}{0.000000\du}}\definecolor{dialinecolor}{rgb}{0.000000, 0.000000, 0.000000}
\pgfsetstrokecolor{dialinecolor}
\pgfsetstrokeopacity{1.000000}
\draw (52.325000\du,23.140000\du)--(52.325000\du,25.040000\du)--(58.330000\du,25.040000\du)--(58.330000\du,23.140000\du)--cycle;
}
\definecolor{dialinecolor}{rgb}{0.000000, 0.000000, 0.000000}
\pgfsetstrokecolor{dialinecolor}
\pgfsetstrokeopacity{1.000000}
\definecolor{diafillcolor}{rgb}{0.000000, 0.000000, 0.000000}
\pgfsetfillcolor{diafillcolor}
\pgfsetfillopacity{1.000000}
\node[anchor=base,inner sep=0pt, outer sep=0pt,color=dialinecolor] at (55.327500\du,24.285000\du){Internal energy};
\pgfsetlinewidth{0.100000\du}
\pgfsetdash{}{0pt}
\pgfsetmiterjoin
{\pgfsetcornersarced{\pgfpoint{0.000000\du}{0.000000\du}}\definecolor{diafillcolor}{rgb}{1.000000, 1.000000, 1.000000}
\pgfsetfillcolor{diafillcolor}
\pgfsetfillopacity{1.000000}
\fill (43.175600\du,23.140000\du)--(43.175600\du,25.040000\du)--(48.890600\du,25.040000\du)--(48.890600\du,23.140000\du)--cycle;
}{\pgfsetcornersarced{\pgfpoint{0.000000\du}{0.000000\du}}\definecolor{dialinecolor}{rgb}{0.000000, 0.000000, 0.000000}
\pgfsetstrokecolor{dialinecolor}
\pgfsetstrokeopacity{1.000000}
\draw (43.175600\du,23.140000\du)--(43.175600\du,25.040000\du)--(48.890600\du,25.040000\du)--(48.890600\du,23.140000\du)--cycle;
}
\definecolor{dialinecolor}{rgb}{0.000000, 0.000000, 0.000000}
\pgfsetstrokecolor{dialinecolor}
\pgfsetstrokeopacity{1.000000}
\definecolor{diafillcolor}{rgb}{0.000000, 0.000000, 0.000000}
\pgfsetfillcolor{diafillcolor}
\pgfsetfillopacity{1.000000}
\node[anchor=base,inner sep=0pt, outer sep=0pt,color=dialinecolor] at (46.033100\du,24.285000\du){Kinetic energy};
\pgfsetlinewidth{0.100000\du}
\pgfsetdash{}{0pt}
\pgfsetbuttcap
{
\definecolor{diafillcolor}{rgb}{0.000000, 0.000000, 0.000000}
\pgfsetfillcolor{diafillcolor}
\pgfsetfillopacity{1.000000}
\pgfsetarrowsstart{stealth}
\pgfsetarrowsend{stealth}
\definecolor{dialinecolor}{rgb}{0.000000, 0.000000, 0.000000}
\pgfsetstrokecolor{dialinecolor}
\pgfsetstrokeopacity{1.000000}
\draw (39.741300\du,24.090000\du)--(43.175600\du,24.090000\du);
}
\pgfsetlinewidth{0.100000\du}
\pgfsetdash{}{0pt}
\pgfsetbuttcap
{
\definecolor{diafillcolor}{rgb}{0.000000, 0.000000, 0.000000}
\pgfsetfillcolor{diafillcolor}
\pgfsetfillopacity{1.000000}
\pgfsetarrowsend{stealth}
\definecolor{dialinecolor}{rgb}{0.000000, 0.000000, 0.000000}
\pgfsetstrokecolor{dialinecolor}
\pgfsetstrokeopacity{1.000000}
\draw (48.940185\du,24.090000\du)--(52.325000\du,24.090000\du);
}
\definecolor{dialinecolor}{rgb}{0.000000, 0.000000, 0.000000}
\pgfsetstrokecolor{dialinecolor}
\pgfsetstrokeopacity{1.000000}
\definecolor{diafillcolor}{rgb}{0.000000, 0.000000, 0.000000}
\pgfsetfillcolor{diafillcolor}
\pgfsetfillopacity{1.000000}
\node[anchor=base,inner sep=0pt, outer sep=0pt,color=dialinecolor] at (41.458500\du,22.471250\du){Work of gravitational};
\definecolor{dialinecolor}{rgb}{0.000000, 0.000000, 0.000000}
\pgfsetstrokecolor{dialinecolor}
\pgfsetstrokeopacity{1.000000}
\definecolor{diafillcolor}{rgb}{0.000000, 0.000000, 0.000000}
\pgfsetfillcolor{diafillcolor}
\pgfsetfillopacity{1.000000}
\node[anchor=base,inner sep=0pt, outer sep=0pt,color=dialinecolor] at (41.458500\du,23.271250\du){ forces};
\definecolor{dialinecolor}{rgb}{0.000000, 0.000000, 0.000000}
\pgfsetstrokecolor{dialinecolor}
\pgfsetstrokeopacity{1.000000}
\definecolor{diafillcolor}{rgb}{0.000000, 0.000000, 0.000000}
\pgfsetfillcolor{diafillcolor}
\pgfsetfillopacity{1.000000}
\node[anchor=base,inner sep=0pt, outer sep=0pt,color=dialinecolor] at (50.607800\du,22.871250\du){Dissipation};
\end{tikzpicture}
  \caption{\label{fig:energy_diagram} Diagram representing energy
    transfers between energy reservoirs.}
\end{figure*}
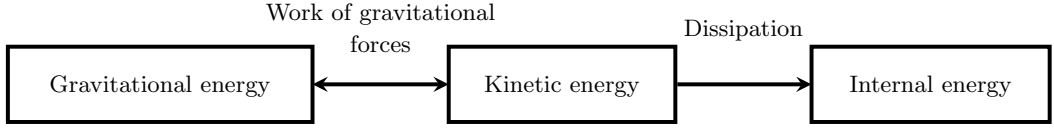

We want to solve Navier-Stokes equations expressing conservation of
mass, balance of momentum and balance of energy, respectively written
as follows
\begin{equation}\label{eq:navier_stokes_gravity_nc}
  \begin{alignedat}{3}
    &\partial_t \rho &&+ \div{\left(\rho \vect{u}\right)} &&= 0,\\
    &\partial_t \left( \rho \vect{u} \right) &&+ \div{\left( \rho
        \vect{u} \otimes \vect{u} + p\tens{I} - \bm{\tau}_{\text{visc}} \right)} &&= \rho \vect{g},\\
    &\partial_t \left( \rho E \right) &&+ \div{\left( \left(\rho E + p\right)\vect{u} -
        \bm{\tau}_{\text{visc}} \vect{u} - \vect{q}_{\text{heat}}\right)} &&= \rho \vect{g} \cdot \vect{u},
  \end{alignedat}
\end{equation}
where $\rho$ is the density, $\vect{u}$ the material velocity, $p$ the
pressure, $\vect{g}$ the external gravitational field,
$\rho E = \rho e + \frac{1}{2} \rho \vect{u}^2$ the density of total
energy with $e$ the specific internal energy, $\vect{q}_{\text{heat}}$
the heat flux and $\bm{\tau}_{\text{visc}}$ the viscous tensor
satisfying
\begin{equation}\label{eq:viscous_tensor}
  \bm{\tau}_{\text{visc}} = \mu \left( \grad{\vect{u}} + \grad{\vect{u}^T} \right) +
  \eta \left(\div{\vect{u}}\right) \tens{I},\\
\end{equation}
where $\mu$ is the dynamic viscosity and $\eta$ the bulk viscosity. We
use $\cdot$ as a scalar product and thus $\div{}$ represents the
divergence operator. In order to close Navier-Stokes
system~\eqref{eq:navier_stokes_gravity_nc} we add constitutive
equations namely a pressure law $p^{\text{EOS}}$~\eqref{eq:eos}, the
Fourier's law~\eqref{eq:fourier_law} and the Stokes
hypothesis~\eqref{eq:stokes_hypothesis}
\begin{subequations}
  \begin{align}
    p &= p^{\text{EOS}} \left( \rho, e \right)\label{eq:eos},\\
    \vect{q}_{\text{heat}} &= - \kappa \grad{T}\label{eq:fourier_law},\\
    \eta &= - \frac{2}{3} \mu\label{eq:stokes_hypothesis}
  \end{align}
\end{subequations}
We recall that the gravitational field is derived from a gravitational
potential $\Phi$ for which $\vect{g} = - \grad{\Phi}$. Dealing with a
constant in time external gravity field, $\partial_t \Phi = 0$ and
using the conservation of mass we
get~\eqref{eq:gravitational_potential_equation}
\begin{equation}
  \label{eq:gravitational_potential_equation}
  \partial_t \left( \rho \Phi \right) + \div{\left(\rho \Phi
      \vect{u}\right)} = \rho \vect{u} \cdot \grad{\Phi}.
\end{equation}
Let us emphasize that in this equation, the gravitational energy
$\rho (\vect{x}, t) \Phi (\vect{x})$ is time dependent only through
the density $\rho (\vect{x}, t)$. Hence the energy
equation~\eqref{eq:gravitational_potential_equation} can be rewritten
in the following conservative form
\begin{equation}\label{eq:conservation_of_energy}
  \partial_t \left( \rho \mathcal{E} \right) + \div{\left( \rho
      \mathcal{E}\vect{u} - \bm{\sigma}_{\text{stress}} \vect{u} -
      \vect{q}_{\text{heat}}\right)} = 0
\end{equation}
where we define
$\rho \mathcal{E} = \rho e + \frac{1}{2} \rho \vect{u}^2 + \rho
\Phi$. Equation~\eqref{eq:conservation_of_energy} expresses the local
conversion between three different energy reservoirs, as depicted in
figure~\ref{fig:energy_diagram}: internal, kinetic and gravitational.
There can be a direct transfer between gravitational energy and
kinetic energy through the work of gravitational forces, from kinetic
energy to internal energy because of the second law of thermodynamics
but no direct transfer between gravitational energy and internal
energy, \refereecorrectionTwo{see also Section 5
  of~\cite{springel_e_2010} and Section 2.2
  of~\cite{marcello_numerical_2012} for a discussion on energy
  conservation for both external and self-gravity cases}.\par{}
Because of this conservation of energy including gravitational energy
we will use the formulation~\eqref{eq:conservation_of_energy} of the
energy equation and we will use the gravitational potential instead of
the usual gravitational field $\vect{g}$. To our knowledge this
approach is quite rare, see~\cite{graham_numerical_1975}
or~\cite{chertock_well-balanced_2018} where they use global fluxes to
have a well-balanced and conservative scheme.\par{}
An important steady state solution of this system for stratified
objects is the hydrostatic balance. The flow is static and the
gravitational force is balanced by the pressure forces, i.e.\
following equation~\eqref{eq:continuous_equilibrium}
\begin{equation}
  \label{eq:continuous_equilibrium}
  \grad{p} = - \rho \grad{\Phi}, \qquad \vect{u} = \vect{0},
\end{equation}
As we mentioned in the introduction, convective flows can be
considered as a perturbation flow of the hydrostatic equilibrium. Thus
this steady state is particularly important in order to study
convection problems in stratified flows.
\section{Numerical scheme}\label{sec:numerical_scheme}
\subsection{Euler system --- Hyperbolic system}
Before going into the derivation of the scheme we introduce the
notations. We define by $\Delta x$ (resp.\ $\Delta y$ and $\Delta z$)
the step along the x-direction (resp.\ the y and z-direction). We note
by $\Delta t$ the time interval between current time $t^n$ and
$t^{n+1}$. We use the notation $q^n_i$ (resp.\ $q^n_{i, j, k}$) to
represent the averaged quantity associated to the field $q$ at time
$t^n$ and in the cell $i$ (resp.\ $i, j, k$) in the one-dimensional
case (resp.\ the three-dimensional case). We use the notation
$q^n_{i+1/2}$ (resp.\ $q^n_{i+1/2, j, k}$) to represent the quantity
associated to the field $q$ at time $t^n$ and at the interface between
cells $i$ and $i+1$ (resp.\ $i, j, k$ and $i+1, j, k$) in the
one-dimensional case (resp.\ the three-dimensional case). Finally we
define the notation ${[q]}_i = q_{i+1/2} - q_{i-1/2}$ in the
one-dimensional case.
\subsubsection{Acoustic-Transport splitting approach}
Following~\cite{chalons_large_2017} we use a splitting strategy that
separates acoustic terms and transport terms and we choose to add the
gravitational source terms to the acoustic part. This way, pressure
gradient can be balanced by the gravity source term.\par{}
However we have another equation compared to the shallow water system
that is the energy equation. As in~\cite{chalons_large_2017}, we want
an isentropic acoustic step for smooth solutions. Thereby we choose to
solve the equation on the gravitational energy,
\begin{equation}\label{eq:navier_stokes_system_c}
  \begin{alignedat}{3}
    &\partial_t \rho &&+ \div{\left(\rho \vect{u}\right)} &&= 0,\\
    &\partial_t \left( \rho \vect{u} \right) &&+ \div{\left(\rho
        \vect{u} \otimes \vect{u} + p\tens{I}\right)} &&= - \rho \grad{\Phi},\\
    &\partial_t \left( \rho \mathcal{E} \right) &&+ \div{\left( \left(\rho
        \mathcal{E} + p\right)\vect{u}\right)} &&= 0,\\
    &\partial_t \left(\rho \Phi\right) &&+ \div{\left( \rho \Phi
        \vect{u} \right)} &&= \rho \vect{u} \cdot \grad{\Phi},
  \end{alignedat}
\end{equation}
\begin{equation*}
  \rho \mathcal{E}= \rho e + \frac{1}{2} \rho \vect{u}^2 + \rho \Phi.
\end{equation*}
However, this leads to a non constant gravitational potential in the
acoustic step whose time variations are compensated in the transport
step in order to have a constant potential in the full step. The
potential is constant in the full step at the continuous level, but
discretization errors with the splitting can lead to a non-constant
discretized potential. Thus we choose to introduce an approximation of
the gravitational called $\Psi \approx \Phi$ and a relaxation
parameter $\lambda$ 
\begin{equation}\label{eq:relaxed_system}
  \begin{alignedat}{3}
    &\partial_t \rho &&+ \div{\left(\rho \vect{u}\right)} &&= 0,\\
    &\partial_t \left( \rho \vect{u} \right) &&+ \div{\left(\rho
        \vect{u} \otimes \vect{u} + p\tens{I}\right)} &&= - \rho \grad{\Phi},\\
    &\partial_t \left( \rho \mathcal{E} \right) &&+ \div{\left( \left(\rho
        \mathcal{E} + p\right)\vect{u}\right)} &&= 0,\\
    &\partial_t \left(\rho \Psi\right) &&+ \div{\left( \rho \Psi
        \vect{u} \right)} &&= \rho \vect{u} \cdot \grad{\Phi} +
    \frac{\rho}{\lambda} \left( \Phi - \Psi \right),
  \end{alignedat}
\end{equation}
\begin{equation*}
  \rho \mathcal{E}= \rho e + \frac{1}{2} \rho \vect{u}^2 + \rho \Psi.
\end{equation*}
We consider the relaxation system~\eqref{eq:relaxed_system} to be an
approximation of the original system~\eqref{eq:navier_stokes_system_c}
that we formally recover in the limit $\lambda \to
0$. System~\eqref{eq:relaxed_system} is solved by first solving the
system in the limit $\lambda \to \infty$ and then in the limit
$\lambda \to 0$ in which $\Psi$ is projected onto $\Phi$, the initial
condition. This way, the evolution of the gravitational potential
$\Psi$, consistent with zero, is forced to be constant. \refereecorrection{
The relaxation technic used here for the gravitational potential is similar
to what is done for pressure relaxation in many approximate Riemann solvers and 
we emphasize that $\Psi$ is just an intermediate used to design the scheme
and can be removed when writing the final scheme (see \ref{sec:overall-algo})
.}\par{}
We now turn to the discretization of the
system~\eqref{eq:relaxed_system} in the limit $\lambda \to
\infty$. Transport phenomena of the form $\vect{u} \cdot \grad{}$ are
separated from the other terms to give two subsystems, first the
acoustic subsystem~\eqref{eq:acoustic_subsystem}
\begin{equation}
  \label{eq:acoustic_subsystem}
  \begin{alignedat}{5}
    &\partial_t \rho &&+ & \rho & &&\div{\vect{u}} &&= 0,\\
    &\partial_t \left( \rho \vect{u} \right) &&+ &\rho \vect{u}& &&\div{\vect{u}} + \grad{p}&&= - \rho \grad{\Phi},\\
    &\partial_t \left( \rho \mathcal{E} \right) &&+ &\rho \mathcal{E}& &&\div{\vect{u}} + \div{\left(p\vect{u}\right)} &&= 0,\\
    &\partial_t \left( \rho \Psi \right) &&+ &\rho \Psi&
    &&\div{\vect{u}} &&= \rho \vect{u} \cdot \grad{\Phi},
  \end{alignedat}
\end{equation}
then the transport subsystem~\eqref{eq:transport_system}
\begin{equation}
  \label{eq:transport_system}
  \begin{alignedat}{3}
    &\partial_t \rho &&+ \vect{u} \cdot \grad{\rho} &&= 0,\\
    &\partial_t \left( \rho \vect{u} \right) &&+ \vect{u} \cdot \grad{\left(\rho \vect{u}\right)} &&= \vect{0},\\
    &\partial_t \left( \rho \mathcal{E} \right) &&+ \vect{u} \cdot \grad{\left(\rho \mathcal{E}\right)} &&= 0,\\
    &\partial_t \left(\rho \Psi\right) &&+ \vect{u} \cdot \grad{\left(\rho \Psi\right)} &&= 0.
  \end{alignedat}
\end{equation}
We now briefly study the eigenstructure of
systems~\eqref{eq:acoustic_subsystem}-\eqref{eq:transport_system}.
Let $\vect{n}$ be any unit normal vector, the acoustic
system~\eqref{eq:acoustic_subsystem} involves seven eigenvalues:
$-c, 0, c$. The fields associated with 0 (resp.\ $\pm c$) are linearly
degenerate (resp.\ genuinely nonlinear), see
Appendix~\ref{appendix:eigenstructure} for more details. The
eigenvalues for transport system~\eqref{eq:transport_system} are given
by $\vect{u} \cdot \vect{n}$. Both
systems~\eqref{eq:acoustic_subsystem}-\eqref{eq:transport_system} are
hyperbolic. \refereecorrection{We emphasize here that the choice of using a
relaxation procedure for the gravitational potential by introducing the equation
on the gravitational potential energy $\rho \Psi$ has been made to obtain this 
simple wave pattern for the splitted Euler system with gravity. (i.e. the same 
pattern as without gravity). Other choices for the relaxation procedure 
(e.g. $\partial_t \Psi = 0$ in both steps) would either lead to the 
introduction of $\vect{u} \cdot \vect{n}$ in the eigenvalues of the acoustic
subsystem or would significantly complexify the relaxation procedure for the 
pressure.}.\par{}
To summarize our numerical procedure, we propose to define a flux
interface by approximating system~\eqref{eq:navier_stokes_system_c}
with a three-step procedure that involves solving the acoustic
system~\eqref{eq:acoustic_subsystem} (acoustic step), the transport
system~\eqref{eq:transport_system} (acoustic step) and finally project
$\Psi$ onto $\Phi$ (relaxation step). We detail each step in the next
sections using the one-dimensional equations.
\subsubsection{Acoustic step}\label{subsub:acoustic_system}
We follow the idea of~\cite{chalons_all-regime_2016} to discretize the
acoustic subsystem. They introduce a pressure relaxation
$\Pi \approx p$, an acoustic impedance $a \approx \rho c$ and a
relaxation parameter $\nu$ to get a fully linearly degenerated
system. It is then written using Lagrangian variables
$(\tau, u, v, \mathcal{E}, \Psi)$ where $u$ represents the normal
velocity component at an interface and $v$ a transverse component. We
also use a mass variable $\mathrm{d} m = \rho(t^n, x) \mathrm{d} x$
where time is frozen at instant $t^n$
\begin{alignat*}{3}
  &\partial_t \tau &&- \partial_m u &&= 0,\\
  &\partial_t u    &&+ \partial_m \Pi &&= - \frac{1}{\tau} \partial_m \Phi,\\
  &\partial_t v    && &&= 0,\\
  &\partial_t \mathcal{E} &&+ \partial_m \left(\Pi u\right) &&= 0,\\
  &\partial_t \Pi &&+ a^2 \partial_m u &&= \frac{1}{\nu} \left( \Pi - p \right),\\
  &\partial_t \Psi && &&= \frac{u}{\tau} \partial_m \Phi,
\end{alignat*}
where
\begin{equation*}
  \mathcal{E} = e + \frac{1}{2} (u^2 + v^2) + \Psi.
\end{equation*}
The discretization of this system is realized with an approximate
Riemann solver that accounts for the source term by means of integral
consistency and composed by three waves $-a, 0, a$,
see~\cite{gallice_solveurs_2002, chalons_large_2013,
  chalons_large_2017}. After the relaxation, in which $\nu \to 0$, it
gives
\begin{alignat*}{2}
  &\widetilde{\tau}_i &&= \tau^n_i + \frac{\Delta t}{\Delta m_i} {\left[ u^* \right]}_i,\\
  &\widetilde{u}_i &&= u^n_i - \frac{\Delta t}{\Delta m_i} {\left[ \Pi^* \right]}_i + \frac{\Delta t}{\Delta m_i} S^n_i,\\
  &\widetilde{v}_i &&= v^n_i,\\
  &\widetilde{\mathcal{E}}_i &&= \mathcal{E}^n_i - \frac{\Delta t}{\Delta m_i} {\left[ \Pi^* u^* \right]}_i,\\
  &\widetilde{\Pi}_i &&= p^{\text{EOS}} \left( \frac{1}{\widetilde{\tau}_i}, \widetilde{e}_i \right),\\
  &\widetilde{\Psi}_i &&= \Psi^n_i - \frac{\Delta t}{\Delta m_i} {(uS)}^n_i,
\end{alignat*}
where
\begin{align*}
  u^*_{i+1/2} &= \frac{1}{2} (u^n_{i+1} + u^n_i) - \frac{1}{2 a} \left( \Pi^n_{i+1} - \Pi^n_i - S^n_{i+1/2}\right),\\
  \Pi^*_{i+1/2} &= \frac{1}{2} \left(\Pi^n_{i+1} + \Pi^n_i\right) - \frac{a^n_{i+1/2}}{2} \left(u^n_{i+1} - u^n_i\right),\\
  a^n_{i+1/2} &\geq \max{\left(\rho^n_i c^n_i, \rho^n_{i+1} c^n_{i+1}\right)},\\
  S^n_i &= \frac{1}{2} \left( S^n_{i+1/2} + S^n_{i-1/2} \right),\\
  {(uS)}^n_i &= \frac{1}{2} (u^*_{i+1/2} S^n_{i+1/2} + u^*_{i-1/2} S^n_{i-1/2}),\\
  S^n_{i+1/2} &= - \frac{1}{2} \left( \frac{1}{\tau^n_i} + \frac{1}{\tau^n_{i+1}} \right) \left( \Phi^n_{i+1} - \Phi^n_i \right).
\end{align*}
and
$a^n_{i+1/2} \geq \max{\left(\rho^n_i c^n_i, \rho^n_{i+1}
    c^n_{i+1}\right)}$ which is the so-called sub-characteristic
condition~\citep[see][]{chalons_large_2013}.\par{}
The update of the conservative variables is then
\begin{alignat*}{2}
  &\widetilde{L}_i \widetilde{\rho}_i &&= \rho^n_i,\\
  &\widetilde{L}_i \widetilde{\left(\rho u\right)}_i &&= {\left(\rho u\right)}^n_i - \frac{\Delta t}{\Delta x} {\left[ \Pi^* \right]}_i + \frac{\Delta t}{\Delta x} S^n_i,\\
  &\widetilde{L}_i \widetilde{\left(\rho v\right)}_i &&= {\left(\rho v\right)}^n_i,\\
  &\widetilde{L}_i \widetilde{\left(\rho \mathcal{E}\right)}_i &&= {\left(\rho \mathcal{E}\right)}^n_i - \frac{\Delta t}{\Delta x} {\left[ \Pi^* u^* \right]}_i,\\
  &\widetilde{L}_i \widetilde{\left(\rho \Psi\right)}_i &&= {\left(\rho \Psi\right)}^n_i - \frac{\Delta t}{\Delta x} {(uS)}^n_i
\end{alignat*}
where
$\widetilde{L}_i = 1 + \frac{\Delta t}{\Delta x} {\left[ u^* \right]}_i$.
\subsubsection{Transport step}\label{subsub:transport_system}
The transport subsystem can be written in the following form, for
$b \in \left\{\rho, \rho u, \rho v, \rho \mathcal{E}, \rho \Psi\right\}$
\begin{equation*}
  \partial_t b + \partial_x \left( b u \right) - b \partial_x u = 0,
\end{equation*}
that is discretized as follows
\begin{equation*}
  b^{n+1}_i = \widetilde{b}_i - \frac{\Delta t}{\Delta x} {\left[ \widetilde{b} u^* \right]}_i + \widetilde{b}_i \frac{\Delta t}{\Delta x} {\left[ u^* \right]}_i.
\end{equation*}
The interface term $\widetilde{b}_{i+1/2}$ is defined by the upwind choice
with respect to the velocity $u^*_{i+1/2}$
\begin{equation*}
  \widetilde{b}_{i+1/2} =
  \begin{cases}
    \widetilde{b}_i \quad \text{if} \quad u^*_{i+1/2} \geq 0\\
    \widetilde{b}_{i+1} \quad \text{if} \quad u^*_{i+1/2} \leq 0\\
  \end{cases}
\end{equation*}
\subsubsection{Relaxation step}
At this stage, the relaxed gravitational potential $\Psi$ still
evolves in time. So we perform the relaxation $\lambda \to 0$ that
boils down to set $\Psi^{n+1}_i = \Phi_i$.
\refereecorrection{
\subsubsection{Overall algorithm}\label{sec:overall-algo}
Gathering the previous steps and intermediate variables, the overall
scheme reads
\begin{equation}
  \label{eq:all_regime_scheme}
  \begin{alignedat}{2}
    &\rho^{n+1}_i &&= \rho^n_i - \frac{\Delta t}{\Delta x} {\left[ \widetilde{\rho} u^* \right]}_i,\\
    &{\left(\rho u\right)}^{n+1}_i &&= {\left(\rho u\right)}^n_i - \frac{\Delta t}{\Delta x} {\left[  \widetilde{\left(\rho u\right)} u^* + \Pi^* \right]}_i + \frac{\Delta t}{\Delta x} S^n_i,\\
    &{\left(\rho v\right)}^{n+1}_i &&= {\left(\rho v\right)}^n_i - \frac{\Delta t}{\Delta x} {\left[  \widetilde{\left(\rho v\right)} u^* \right]}_i,\\
    &{\left(\rho \mathcal{E}\right)}^{n+1}_i &&= {\left(\rho \mathcal{E}\right)}^n_i - \frac{\Delta t}{\Delta x} {\left[ \left(\widetilde{\left(\rho \mathcal{E}\right)} + \Pi^*\right) u^* \right]}_i
  \end{alignedat}
\end{equation}
It may also be expressed as a \refereecorrectionTwo{first-order}
classic finite-volume scheme involving flux terms for the conservative
part for energy $\rho E = \rho e + \frac{1}{2} \rho u^2$ and source
terms for gravity
\begin{equation}
  \begin{alignedat}{2}
      &\rho^{n+1}_i &&= \rho^n_i - \frac{\Delta t}{\Delta x} {\left[ \widetilde{\rho} u^* \right]}_i,\\
      &{\left(\rho u\right)}^{n+1}_i &&= {\left(\rho u\right)}^n_i - \frac{\Delta t}{\Delta x} {\left[  \widetilde{\left(\rho u\right)} u^* + \Pi^* \right]}_i - \Delta t {\left\{ \rho \partial_x \Phi \right\}}_i,\\
      &{\left(\rho v\right)}^{n+1}_i &&= {\left(\rho v\right)}^n_i - \frac{\Delta t}{\Delta x} {\left[  \widetilde{\left(\rho v\right)} u^* \right]}_i,\\
      &{\left(\rho E\right)}^{n+1}_i &&= {\left(\rho E\right)}^n_i - \frac{\Delta t}{\Delta x} {\left[ \left(\widetilde{\left(\rho E\right)}^{NG} + \Pi^*\right) u^* \right]}_i - \Delta t {\left\{ \rho u \partial_x \Phi \right\}}_i,
    \end{alignedat}
\end{equation}
where
\begin{align*}
  \Delta x {\left\{ \rho u \partial_x \Phi \right\}}_i &= {\left[
  \widetilde{\rho} u^* \Phi \right]}_i - {\left[\widetilde{\rho}
  u^*\right]}_i \Phi_i,\\
  \Delta x {\left\{ \rho \partial_x \Phi \right\}}_i &= - S^n_i,\\
  \widetilde{\left(\rho E\right)}^{NG}_i &= {\left(\rho E\right)}^n_i - \frac{\Delta t}{\Delta x} {\left[ \Pi^* u^* \right]}_i.
\end{align*}
We emphasize that both formulations are equivalent and conservative
for the energy $\rho \mathcal{E}$.} A non-conservative energy approach
is also detailed in Appendix~\ref{appendix:non-conservative}.\par{}
We can notice that in the case of a constant gravitational potential,
we recover the original scheme derived
in~\cite{chalons_all-regime_2016}.
\subsubsection{On the low-Mach correction}
As for the scheme of~\cite{chalons_all-regime_2016} and as explained
in~\cite{dellacherie_analysis_2010}, the numerical scheme defined
by~\eqref{eq:all_regime_scheme} poorly performs in the low Mach regime
due to truncature error of magnitude $\frac{\Delta x}{\mathrm{Ma}}$
that comes from the term $\Pi^*_{i+1/2}$. To tackle this issue,
following~\cite{chalons_all-regime_2016} we modify the upwinding part
of $\Pi^*_{i+1/2}$ thanks to an extra parameter $\theta_{i+1/2}$ by
setting
\begin{equation*}
  \Pi^*_{i+1/2} = \frac{1}{2} \left(\Pi^n_{i+1} + \Pi^n_i\right) - \frac{a^n_{i+1/2} \theta_{i+1/2}}{2} \left(u^n_{i+1} - u^n_i\right),
\end{equation*}
\begin{equation}\label{eq:low_mach_correction}
  \begin{aligned}
    \theta_{i+1/2} &= \min{\left( \mathrm{Ma}_{i+1/2}, 1\right)},\\
    \mathrm{Ma}_{i+1/2} &= \frac{\vert u^*_{i+1/2}
      \vert}{\max{\left(c^n_i, c^n_{i+1}\right)}}.
  \end{aligned}
\end{equation}
\refereecorrection{Using a truncation analysis in dimensionless form
  it can be shown that this correction acts like a rescaling of the
  numerical diffusion induced by the pressure
  discretization~\citep[see][]{chalons_all-regime_2016}.}\par{}
As we can see, the low-Mach correction does not directly come from the
derivation of the numerical scheme~\ref{eq:all_regime_scheme}. Some
ongoing works are trying to derive directly all-Mach schemes using
more sophisticated relaxation
schemes~\citep[see][]{bouchut_entropy_2017-1}.
\subsubsection{On the well-balanced property}
A numerical scheme is said to be well-balanced for equilibrium states
satisfying equation~\eqref{eq:continuous_equilibrium}, if it exists a
discrete counterpart of equation~\eqref{eq:continuous_equilibrium} in
which solutions are preserved by the numerical scheme.\par{}
The discrete counterpart of equation~\eqref{eq:continuous_equilibrium}
for scheme~\eqref{eq:all_regime_scheme} is given by
\begin{gather}
  \begin{gathered}\label{eq:well_balanced_formula}
    u^n_i = 0, \quad v^n_i = 0,\\
    \Pi^n_{i+1} - \Pi^n_i = - \frac{1}{2} \left( \rho^n_i + \rho^n_{i+1}
    \right) \left( \Phi_{i+1} - \Phi_i \right),
  \end{gathered}
\end{gather}
Let us now verify that we have obtained a well-balanced scheme. If at
time $t^n$, for some density profile the initial state reads as
in~\eqref{eq:well_balanced_formula} then fluxes from the acoustic step
reduce to
\begin{gather*}
  u^*_{i-1/2} = u^*_{i+1/2} = 0\\
  {[\Pi^*]}_i = \frac{1}{2} \left( \Pi^n_{i+1} - \Pi^n_i \right) + \frac{1}{2} \left( \Pi^n_i - \Pi^n_{j-1} \right) + S^n_i.
\end{gather*}
Then we have for the acoustic step
\begin{gather*}
  \widetilde{u}_i = u^n_i, \quad \widetilde{v}_i = v^n_i,\\
  \widetilde{\rho}_i = \rho^n_i, \quad  \widetilde{\mathcal{E}}_i = \mathcal{E}^n_i.
\end{gather*}
Finally, because $u^*_{i+1/2}$ vanishes, transport step is trivial and
the initial state remains unchanged. \refereecorrection{Once we have made the 
appropriate choice for the discretization of the gravitational source term in 
the acoustic step, the well-balanced property is automatically verified without 
the need to introduce an other algorithmic correction.}
\subsection{Dissipative fluxes --- Parabolic
  system}
We now turn to the discretization of dissipative
fluxes~\eqref{eq:viscous_tensor}-\eqref{eq:fourier_law}. They are
discretized using first order discrete fluxes
\begin{align*}
  {\left[ \div{\vect{f}^{dissipative}} \right]}_{i, j, k} = \phantom{+} &\frac{(f_{x, i+1/2, j, k} - f_{x, i-1/2, j, k})}{\Delta x}\\
  + &\frac{(f_{y, i, j+1/2, k} - f_{y, i, j-1/2, k})}{\Delta y}\\
  + &\frac{(f_{z, i, j, k+1/2} - f_{z, i, j, k-1/2})}{\Delta z}
\end{align*}
where $\vect{f}^{dissipative}$ is either the heat flux
$\vect{q}_{\text{heat}}$ or the viscous flux
$\bm{\tau}_{\text{visc}}$. In the case of the heat flux we have
\begin{gather*}
  q_{x, i+1/2, j, k} = - \kappa \frac{\left(T_{i+1, j, k} - T_{i  , j, k}\right)}{\Delta x}\\
  q_{y, i, j+1/2, k} = - \kappa \frac{\left(T_{i, j+1, k} - T_{i, j  , k}\right)}{\Delta y}\\
  q_{z, i, j, k+1/2} = - \kappa \frac{\left(T_{i, j, k+1} - T_{i, j, k  }\right)}{\Delta z}
\end{gather*}
With the addition of the viscous terms and the heat flux, this
all-regime well-balanced scheme is now well-suited for the study of
convection problems in highly stratified flows in both low Mach and
high Mach regimes. Before showing validating numerical tests, we
present some specificities about the numerical implementation and
parallelization used in this work.
\section{Implementation and parallelization}\label{sec:implementation}
\begin{figure}
  \centering
  \includegraphics{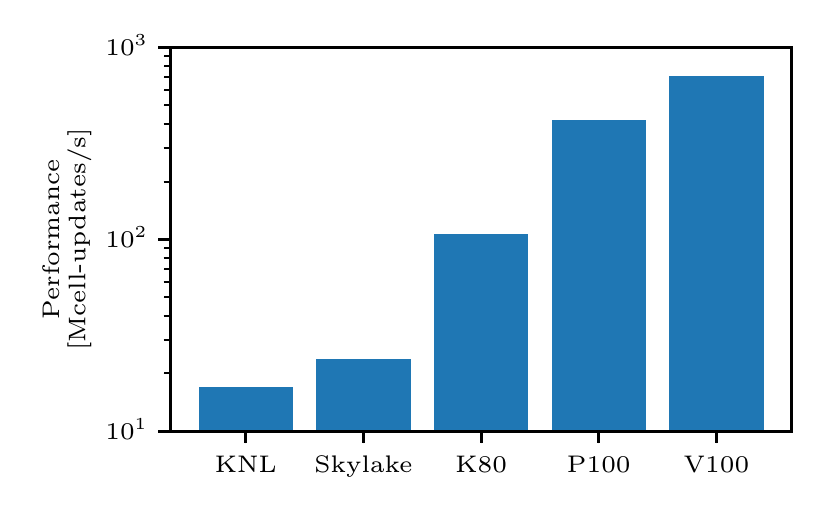}
  \caption{\label{fig:performance_comparison} Comparison of
    performance on different architectures: Intel KNL, Intel Skylake
    (one socket), \refereecorrection{\uppercase{Nvidia} K80,
      \uppercase{Nvidia} P100 and \uppercase{Nvidia} V100}. Measures
    on Intel KNL and Intel Skylake were performed on Joliot-Curie's
    supercomputer at TGCC using the same code. In our case we obtain
    better results with the Intel Skylake than the Intel KNL due to a
    lack of vectorization. Going to a GP-GPU we have a speed-up around
    five with a \refereecorrection{\uppercase{Nvidia}} K80 compared to
    multi-core architecture \refereecorrection{and seven between
      \uppercase{Nvidia} K80 and V100}.}
\end{figure}
In this section we describe the implementation of the scheme using
Kokkos library. We begin by giving a brief overview of the Kokkos
library.
\subsection{Exascale computing}
To reach the exascale, the distributed memory model is not sufficient
to take advantage of all the computing power of new
architectures. There are mainly two reasons for this. First, nodes of
supercomputers tend to grow more and more and hence are more suited to
a shared memory model~\citep{sunderland_overview_2016}. Secondly,
nodes tend be more and more heterogeneous by using multi-core,
many-core and/or accelerators like GP-GPUs. So it means that even if
shared memory is exposed, it needs to be handled differently from one
architecture to another. For example we can think of OpenMP or C++11
threads for multi-core and many-core processors, and CUDA or OpenACC
for GP-GPUs.\par{}
Moreover this architecture heterogeneity raises a performance
portability issue. Currently, many HPC codes are optimized for some
specific architectures to get the maximum computing power. However
this optimization process couples the numerical scheme to its
implementation details like the memory management, the loop ordering,
cache blocking and so on. Hence running a code on a different
architecture results in bad performance.\par{}
We propose to use the recent C++ library
Kokkos~\citep[see][]{carter_edwards_kokkos:_2014} that implements a
new shared memory model. Using abstract concepts such as execution
spaces (where a function is executed), data spaces (where data
resides) and execution policies (how the function is executed) the
library is able to efficiently take advantage of multi-core many-core
processors and GP-GPUs. This way the portability relies on the library
and no more on the numerical code.
\subsection{Implementation}
Following the work of~\cite{kestener_implementing_2017}, the code is
then organized with computation kernels:
\begin{itemize}
  \item Acoustic and transport kernels,
  \item Viscous and heat diffusion operator kernels,
  \item Conservative variables to primitive variables kernel,
  \item Time step kernel.
\end{itemize}
Each kernel is a C++ functor. They are given to Kokkos through the
function \lstinline[language=C++]{Kokkos::parallel_for}. Internally,
depending on the device chosen at compile-time, it hides a
parallelized one-dimensional loop where the current index is given as
an argument to the functor. This index is then interpreted as a cell
index in the domain.\par{}
Kokkos only deals with shared memory systems. We use the Message
Passing Interface (MPI) programming model with a regular domain
decomposition to take advantage of distributed memory machines across
multiples computing nodes. Kokkos is then used as a shared memory
programming model inside each node. These domains are endowed with
ghost zones which are used to both implement physical boundary
conditions and to contain values from neighbour
domains. Communications are handled through the ghost cell
pattern~\citep[see][]{kjolstad_ghost_2010}. Thus for a given direction
X, Y (or Z) and a given side, left or right, one MPI process sends
data from its domain to its neighbour's ghost zone and receives data
into its own ghost zone.
\subsection{Performance results}
Thanks to Kokkos, we were able to use \textbf{the same code} on
different architectures like Intel Skylake, Intel Knights Landing
(KNL) and \refereecorrection{\uppercase{Nvidia}} GP-GPUs
\refereecorrection{(K80, P100, V100)}\@. We measured performance on
the Intel Skylake and the Intel KNL partition of the Joliot Curie
machine at TGCC\@. Figure~\ref{fig:performance_comparison} shows the
results. We see that the Kokkos library is able to provide good
performance on the different tested architectures. Nevertheless, even
if the peak performance of the Intel KNL architecture is higher than
the Intel Skylake one we have better performance on the Intel Skylake
architecture. We also notice the important speed-up (around five)
between the Intel Skylake architecture and the
\refereecorrection{\uppercase{Nvidia} V100} GP-GPU.\par{}
\begin{figure}
  \centering
  \includegraphics{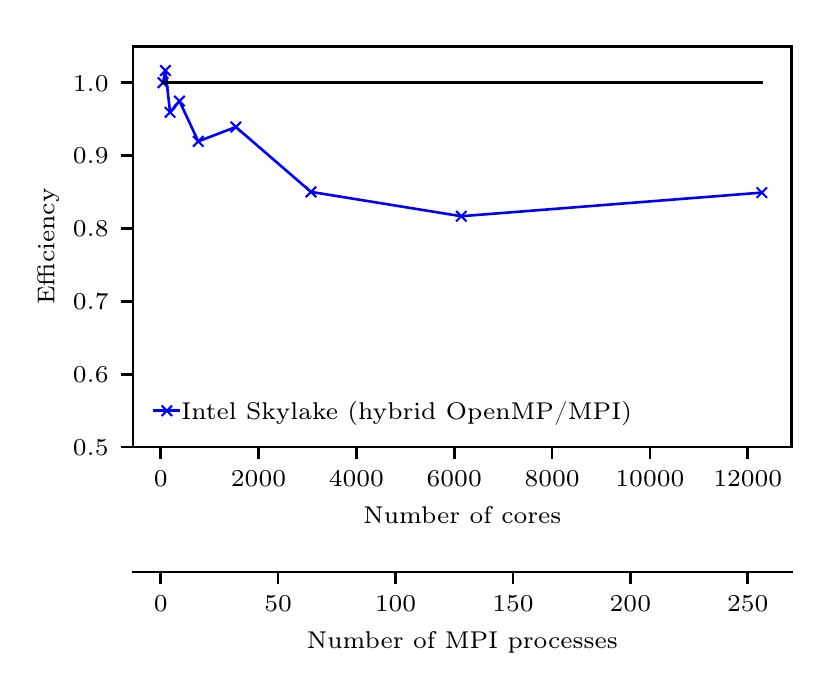}
  \caption{\label{fig:scaling_skylake} Weak scaling results obtained
    on Joliot Curie's Intel Skylake partition at TGCC\@. We use a
    hybrid MPI-OpenMP configuration in which one MPI task is bound to
    a socket. Simulations run for 1000 time steps and each MPI process
    treats $128^3$ cells. We see that the efficiency reaches a plateau
    of 85\%.}
\end{figure}
Figure~\ref{fig:scaling_skylake} shows a weak scaling test performed
with a hybrid configuration OpenMP/MPI\@. We went up to 512 MPI
processes, one MPI process per Intel Skylake socket to avoid NUMA
effects. It results in a total of 12288 cores at 512 MPI
processes. Each MPI process is getting a piece of the whole domain of
$128^3$, so a domain of $44^3$ per core. We can see that we obtain a
plateau of 85\% of maximum performance from 128 MPI processes.\par{}
The performances obtained with the use of the
\refereecorrection{Kokkos} library are encouraging for the study of
convection problems with the \refereecorrection{ARK} code on massively
parallel present and future architectures. In the next section, we use
several numerical tests to show that the numerical scheme used in the
ARK code is indeed very well suited for the study of convection.
\section{Numerical results}\label{sec:numerical_results}
In this section we specialize the equation of state~\ref{eq:eos}. We
will use an ideal gas satisfying
\begin{equation*}
  p^{\text{EOS}}\left( \rho, e \right) = \left( \gamma - 1 \right) \rho e
\end{equation*}
where $\gamma$ is the adiabatic index of the gas. The speed of sound
satifies the following simple relation
\begin{equation*}
  c^2 = \gamma \frac{p}{\rho}
\end{equation*}
We emphasize that it is possible to use a different equation of state
with the all-regime well-balanced numerical scheme.
Moreover we consider two versions of the all-regime scheme depending
on the low-Mach correction. We will refer to the disabled low-Mach
correction scheme when $\theta = 1$ and to the enabled one when
$\theta$ follows equation~\ref{eq:low_mach_correction}.\par{}
We will test different properties of the scheme with different test
cases: wave speeds with the Sod test (no gravity), low-Mach accuracy
with the Gresho vortex test (no gravity), hydrostatic balance with the
test of an atmosphere at rest and out of equilibrium behavior with the
Rayleigh-Taylor test. We then use the ARK code for the study of
Rayleigh-B\'{e}nard convection.
\subsection{Shock tube test}
The Sod shock tube~\citep{sod_survey_1978} is a classical test for
compressible solvers. It tests the ability of the solver to have
correct wave speeds and its numerical diffusion near
discontinuities.\par
\begin{figure}
  \centering
  \includegraphics{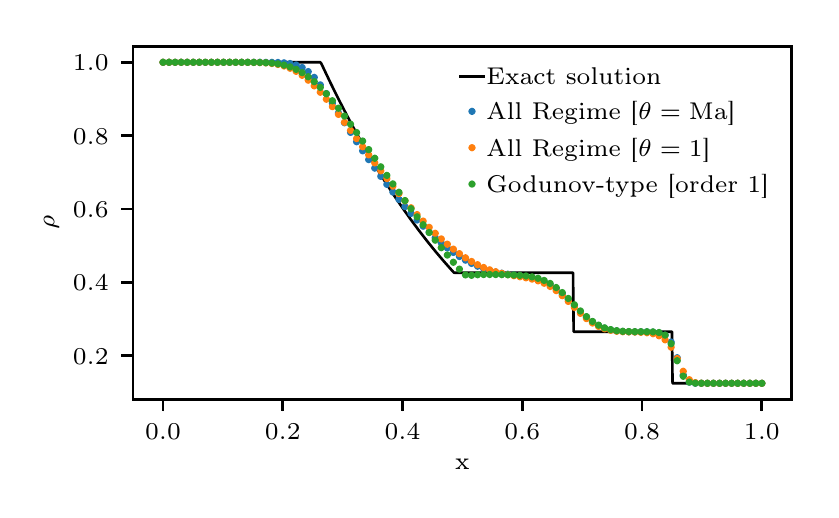}
  \caption{\label{fig:sod_test_case} Sod's test case
    simulations. Figure shows a snapshot of the density profile $\rho$
    for the All-Regime scheme, with and without the low-Mach
    correction, a first order Godunov-type scheme (HLLC) and the exact
    solution. Spatial resolution is nx = 100. We see that the
    All-Regime scheme gives results close to the Godunov-type scheme
    around discontinuities but is more diffusive in the rarefaction
    wave.}
\end{figure}

{}
The computational domain is the interval [0,1], the initial condition
is defined by
\begin{equation*}
  (\rho, p, u) =
  \begin{cases}
    (1, 1, 0) & \quad \text{if} \quad x < 0.5,\\
    (0.125, 0.1, 0) & \quad \text{if} \quad x \geq 0.5.
  \end{cases}
\end{equation*}
Results are shown in figure~\ref{fig:sod_test_case} for simulations
with nx = 100.
First we can observe that the solver is as good as a first order
Godunov-type scheme with a HLLC approximate Riemann solver around the
contact discontinuity and the shock. However the rarefaction wave is a
bit more diffused. We also notice that the low-Mach correction does
not influence the behavior of the scheme for this test case. However
we want to stress out some instability near discontinuities, as shown
in~\cite{chalons_all-regime_2016}. This can also be seen in a double
shock waves test case.
\subsection{Gresho vortex test case}
The Gresho vortex~(\cite{gresho_theory_1990, miczek_new_2015}) is a
test case that has already been used to test numerical schemes in the
low Mach regime. It is a two dimensional stationary test case that can
be parameterized by the maximum value of the Mach number. It is thus
well-suited to study the behavior of the scheme in the low Mach
regime. We recall that the test case is defined using polar
coordinates $(r, \theta)$ defined with respect to the center of the
vortex as follows
\begin{gather*}
  \rho = \rho_0,\\
  \left(u_r, u_{\theta}\right) =
                                 \begin{cases}
                                   \left(0, 5r\right) & 0 \leq r < 0.2, \\
                                   \left(0, 2-5r\right), & 0.2 \leq r < 0.4, \\
                                   \left(0, 0\right) , & 0.4 \leq r
                                 \end{cases}\\
  p =
      \begin{cases}
        p_0 + 12.5 r^2, & 0 \leq r < 0.2, \\
        p_0 + 12.5 r^2 + 4 - 20r + 4 \ln (5r), & 0.2 \leq r < 0.4, \\
        p_0 -2 + 4 \ln 2, & 0.4 \leq r.
      \end{cases}
\end{gather*}
\begin{figure*}
  \centering
  \includegraphics{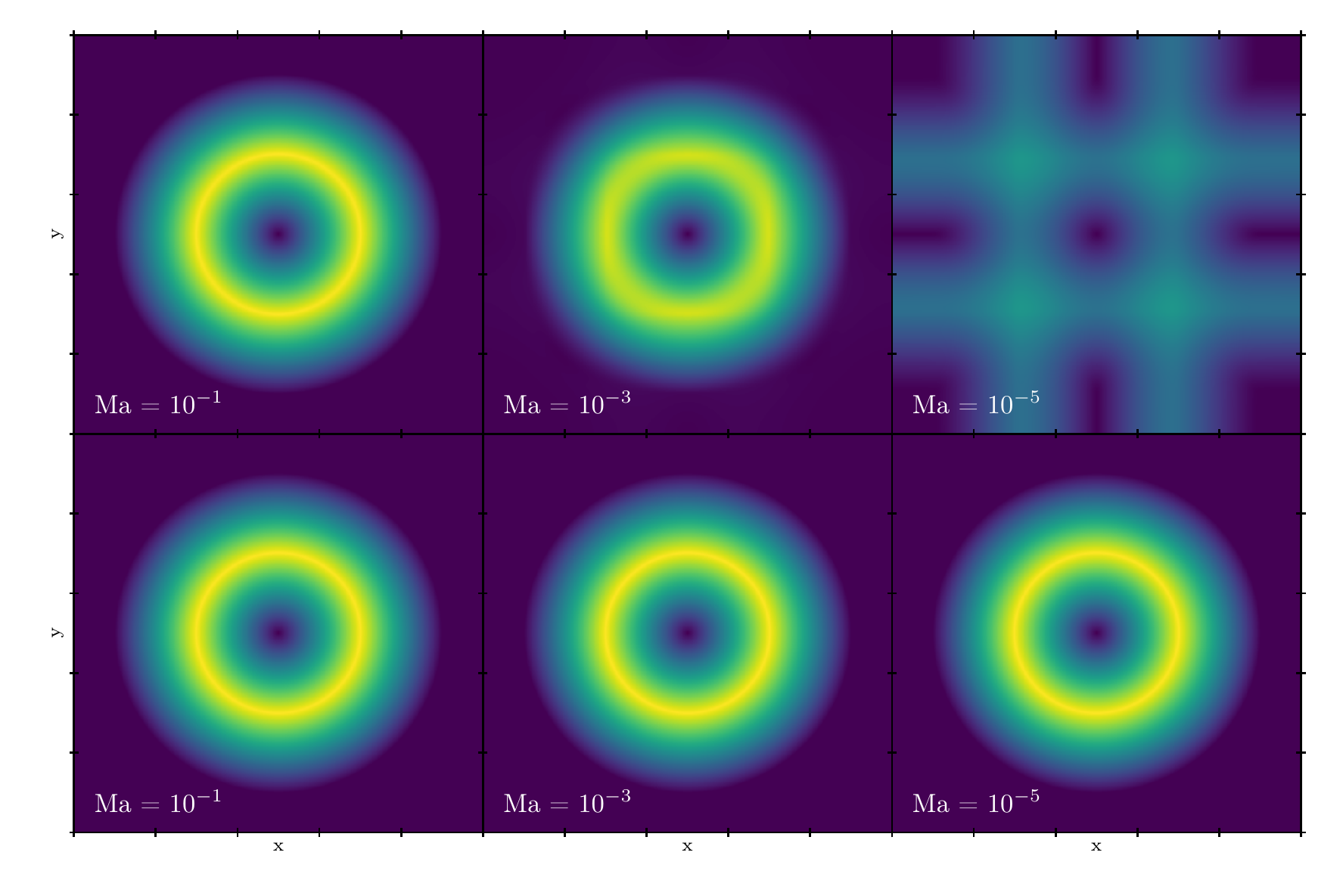}
  \caption{\label{fig:gresho} Gresho vortex simulations. Snapshots of
    the magnitude of the velocity field at time $t_f = 10^{-3}$, for a
    resolution of $512^2$ and for different Mach numbers. First line
    shows results where the low-Mach correction is disabled and second
    line where it is enabled. We see that without the low-Mach
    correction the scheme fails at simulating low-Mach flows.}
\end{figure*}
where $p_0$ satisfies $p_0 = \frac{1}{\gamma \mathrm{Ma}^2}$. In this
case $\mathrm{Ma}$ is a parameter and $\gamma$ is the adiabatic index
of the ideal gas. The velocity is normalized so a particle placed at
the peak of velocity ($u = 1.0$ at location $r = 0.2$) make a full
rotation in $\Delta t = \frac{2}{5} \pi \approx 1.26$.\par{}
We ran a serie of simulations with different solvers where we explored
parameter space nx and Ma from 32 to 2048 and from 1.0 to
$1.0 \times 10^{-5}$ respectively. Final time is set to
$t_f = 1.0 \times 10^{-3}$, which has been chosen sufficiently small
such that the error doesn't saturate.\par
\begin{figure}[ht]
  \centering
  \includegraphics{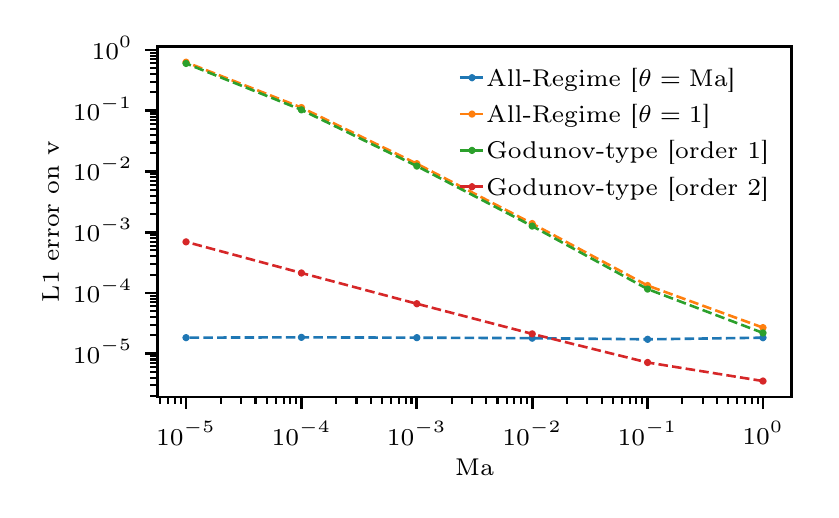}
  \caption{\label{fig:v_vs_Ma_nx_2048} Gresho vortex
    simulations. $L^1$ error on the velocity in function of the Mach
    number at a fixed number of points of nx = 2048}
\end{figure}

\begin{figure}[ht]
  \centering
  \includegraphics{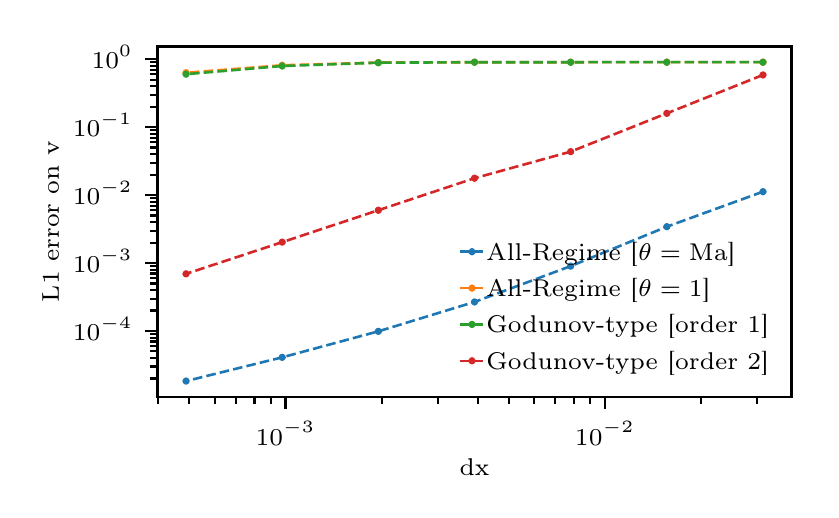}
  \caption{\label{fig:v_vs_dx_Ma_5} Gresho vortex simulations. $L^1$
    error on the velocity in function of the spatial resolution, at a
    fixed Mach number of $\mathrm{Ma} = 10^{-5}$.}
\end{figure}

{}
Figure~\ref{fig:gresho} shows snapshots of the the magnitude of the
velocity field at the final time and at resolution $512^2$. We see
that when the Mach number decreases the velocity field becomes more
and more degraded when the low-Mach correction is disabled. At
$\mathrm{Ma} = 10^{-5}$, the vortex has completely disappeared.
Figures~\ref{fig:v_vs_Ma_nx_2048} and~\ref{fig:v_vs_dx_Ma_5} show more
quantitative results where we show absolute $L^1$ error on velocity in
function of the Mach number Ma and the spatial resolution dx
respectively. Figure~\ref{fig:v_vs_Ma_nx_2048} shows that $L^1$ error
on velocity depends on the Mach number. More precisely we measure a
slope of -1 on schemes or order 1 and a slope of -0.5 on scheme of
order 2. On the other hand the low Mach correction of the all-regime
scheme gives a uniform error with respect to the Mach number.
Figure~\ref{fig:v_vs_dx_Ma_5} shows convergence curves at
$\mathrm{Ma} = 1.0 \times 10^{-3}$. We see that both Godunov-type and
all-regime without the low Mach correction converge at order 1 as
expected. Nevertheless Godunov-type with Muscl-Hancock reconstruction
converges only at order 1.5. It may be due to the lack of regularity
of the velocity field as it can be observed in the case of a contact
discontinuity~\citep[see][]{springel_e_2010}. All-Regime scheme shows
two different behaviors, at first it converges at order 1.5 then
around $\mathrm{nx} = 1024$ the slope changes and it converges at
order 1.2. We assume that at higher resolution we would recover order
1. We see that at low Mach number the precision, independently of the
order, is better than the one of a Godunov-type scheme.
\subsection{Well-balanced test case}
The well-balanced test case is a simple isothermal column of
atmosphere at equilibrium. This column of atmosphere is in a stable
equilibrium state. The test allows us to measure the ability of the
scheme to preserve this equilibrium. After normalization, it is given
by
\begin{gather*}
  p (z) = \rho (z) = e^{-z}\\
  T = 1
\end{gather*}
which is the solution of the following system
\begin{gather*}
  \frac{\mathrm{d} p}{\mathrm{d} z} = - \rho \frac{\mathrm{d} \Phi}{\mathrm{d} z}\\
  T = 1\\
  p = \rho T
\end{gather*}
We take advantage of the formula~\eqref{eq:well_balanced_formula} and we
initialize the test case with the following formula
\begin{gather*}
  \frac{p_{i+1} - p_i}{\Delta z} = - \frac{\rho_i + \rho_{i+1}}{2} \frac{\Phi_{i+1} -\Phi_i}{\Delta z}\\
  T_i = 1\\
  p_i = \rho_i T_i
\end{gather*}
The computational domain used is the interval $[0, 3]$. Results are
displayed in table~\ref{table:well-balanced_results} at time $t=10$,
more than three times the sound crossing time in the box. We see that
we stay near \refereecorrection{machine precision} at the end of the
simulation. We see a shift of two orders of magnitude in the error
when using the low-Mach correction. \refereecorrection{The reason of this shift
is not entirely clear and is difficult to interpret as it involves truncature
errors. Looking at the spatial pattern of the error in the simulation, it does
seem to come from the boundary conditions (extrapolation of the hydrostatic 
balance for pressure and density and reflexive conditions for the velocity) 
with the use of the low-Mach correction. A more appropriate boundary condition
might remove this shift in the error (which is in any case sufficiently small
and stable to allow the use of controlled seeded perturbations).}
\begin{table}
  \caption{Isothermal atmospheres at rest. Table shows for different
    spatial resolutions the maximum velocity in the domain. We see
    that the velocity is maintained \refereecorrection{around zero up
      to the machine precision}, thus illustrating the well-balanced
    property.~\label{table:well-balanced_results}}
  \begin{tabular}{ccc}
    Number of cells & velocity ($\theta=1$) & velocity ($\theta=Ma$)\\
    128& $2.9 10^{-15}$ & $1.4 10^{-13}$\\
    256& $8.1 10^{-15}$ & $5.7 10^{-13}$\\
    512& $1.5 10^{-14}$ & $1.1 10^{-12}$\\
    1024& $2.2 10^{-14}$ & $2.2 10^{-12}$\\
    2048& $4.7 10^{-14}$ & $1.6 10^{-12}$\\
    4096& $1.1 10^{-13}$ & $4.0 10^{-12}$\\
  \end{tabular}
\end{table}
\subsection{Rayleigh-Taylor instability test
  case}
The Rayleigh-Taylor test case is a two dimensional test case where two
fluids of different densities are superposed and are at
equilibrium. The denser one is on top. A small perturbation is
introduced to break equilibrium.\par{}
The full setup is as follow, for a domain
$\left[ -0.25, 0.25 \right] \times \left[ -0.75, 0.75 \right]$:
\begin{align*}
  \rho \left( x, y \right) &=
         \begin{cases}
           1 & \text{for } y < 0 \\
           2 & \text{for } y >= 0
         \end{cases} \\
  p \left( x, y \right) &= \rho g y \\
  u \left( x, y \right) &= 0 \\
  v \left( x, y \right) &= \frac{C}{4} \left( 1 + \cos \left( \frac{2 \pi x}{L_x} \right) \right)
                          \left( 1 + \cos \left( \frac{2 \pi y}{L_y} \right) \right)
\end{align*}
Where $C = 0.01$ is the magnitude of the velocity perturbation,
$L_x = 0.5$ and $L_y = 1.5$ are the size of the domain in each
direction. We do not need to use the well-balanced
formula~\eqref{eq:well_balanced_formula}, the equilibrium is preserved
in the case $A=0$.

\begin{figure*}
  \centering
  \includegraphics{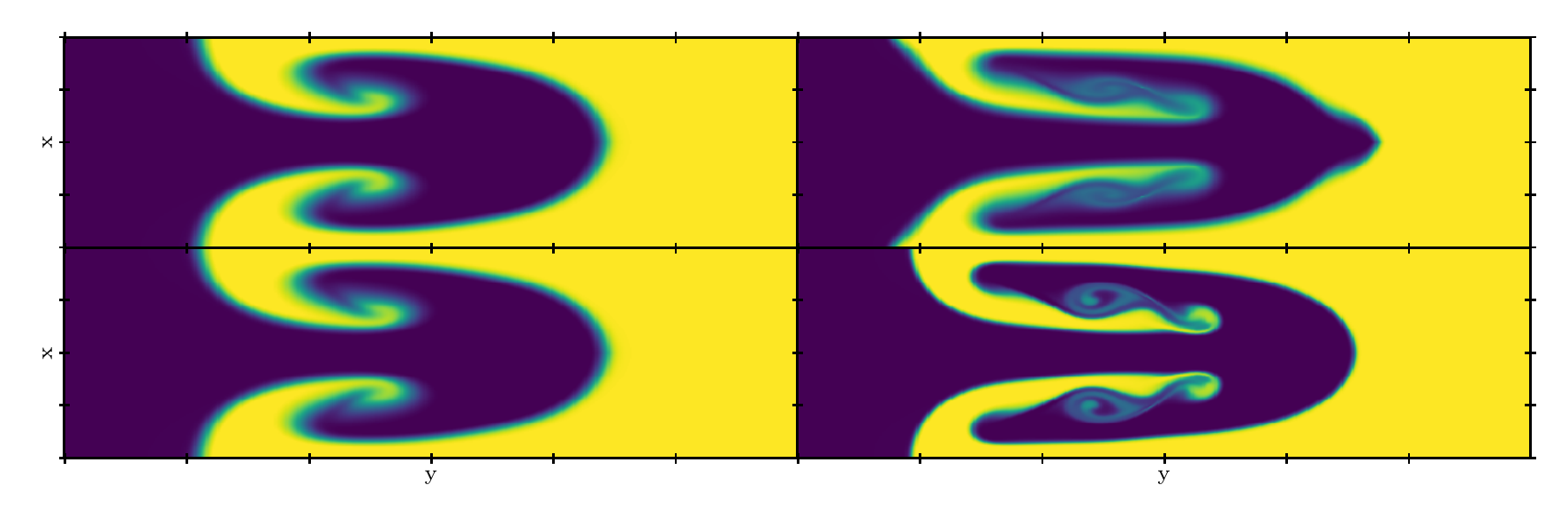}
  \caption{\label{fig:rayleigh_taylor} Rayleigh-Taylor
    simulations. Figure shows snapshots of density, one in purple and
    two in yellow at time $t=12.4$ and for a resolution of
    $200 \times 600$. First line show results with the the all-regime
    scheme, where on the left the low-Mach correction is disabled and
    is enabled on the right. Second line shows results with a
    Godunov-type scheme, on the left it is first order, on the right
    it is second order using a Muscl-Hancock scheme. We see that with
    the low-Mach correction we recover features only present at second
    order for a standard Godunov-type scheme.}
\end{figure*}

Figure~\ref{fig:rayleigh_taylor} shows two simulations of the
Rayleigh-Taylor test case, one with the low Mach correction and the
other without it ($\theta^n_{i+1/2} = 1$). Both simulations are at the
same time $t=12.4$ and the same resolution $200 \times 600$. The
yellow part is at density $2$ and the purple is at density $1$. We see
that we recover the classical linear growing mode. Moreover the
simulation with the low Mach correction is able to capture secondary
instabilities in the non linear regime. They are closer to the second
order Godunov-type simulation than the order one. However the low Mach
correction does not help on the interface diffusion between the two
mediums. It also shows a peak that is not present without the
correction at the same resolution. This spurious behavior is therefore
caused by the low Mach correction that removes some numerical
diffusion in the scheme. \refereecorrection{ By looking at higher
  resolutions, we identify that this peak is a grid-seeded secondary
  RT unstable mode that appears at the top of the large scale seeded
  mode. This type of secondary modes are not unexpected and can be
  seen for example in Fig. 9 of~\cite{almgren_castro:_2010}.} This
peak disappears with the addition of some physical viscosity in the
simulation.
\subsection{Rayleigh-B\'{e}nard instability test
  case}
This last test case is about compressible convection simulations both
in 2D and 3D. In this test case there are different important
parameters. First, from stability analysis we know that the Rayleigh
number $\mathrm{Ra}$ is an important non-dimensional number. Beyond a
threshold, called the critical Rayleigh $\mathrm{Ra}_c$, the
convection process starts and efficiently transports the
heat~\citep[see Figures 1 and 3
in][]{hurlburt_two-dimensional_1984}. Below this threshold, diffusion
processes are sufficient to transport heat and no material
displacement is necessary. Then another important parameter is the
density stratification $\chi$ which the ratio between the density at
the bottom of the domain and the density at the top. In the highly
stratified case, study of convection becomes more difficult as there
is not a unique Rayleigh number but more a whole range of values
extending on the scale height. Notice that when $\chi \to 1$ we
recover the Boussinesq-like situation. Finally the last parameter is
the polytropic index $m$ which is a measure of how close is the
initial temperature gradient from the adiabatic gradient. One can show
that the Schwarzschild criterion writes
\begin{equation*}
  m + 1 < \frac{\gamma}{\gamma-1} = 2.5, \quad \gamma = \frac{5}{3}
\end{equation*}
The initial setup is inspired
from~\cite{hurlburt_two-dimensional_1984,
  toomre_three-dimensional_1990}. Following their notation, the
initial state is given by a polytropic profile of polytropic index $m$
\begin{equation*}
  T = z, \quad \rho = z^m, \quad p = z^{m+1}
\end{equation*}
where $z$ is the vertical variable. It is initialized using to the
recursive formula~\eqref{eq:well_balanced_formula}. So we begin with a
hydrostatic equilibrium that we destabilize whether with a velocity
mode perturbation or with a temperature random perturbation.\par{}
\subsubsection{2D case}
\begin{figure}
  \centering
  \includegraphics{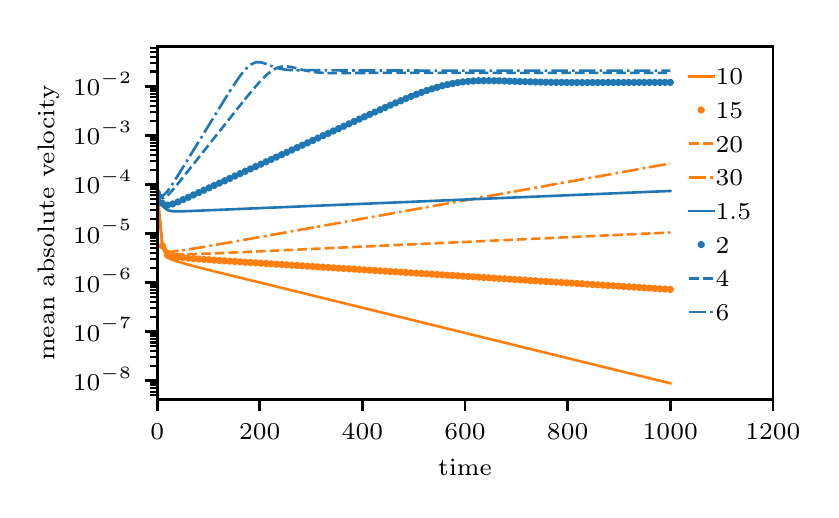}
  \caption{\label{fig:growth_rates} Rayleigh-B\'{e}nard instability
    simulations in 2D. Figure shows the time evolution of the mean
    absolute velocity for different ratios of Rayleigh number over
    critical Rayleigh number (see legend). Blue points show the case
    where the low-Mach correction is enabled and orange ones where it
    is disabled. We observe that when the low-Mach correction is
    enabled the onset of convection is closer to the expected critical
    Rayleigh number.}
\end{figure}

We begin with 2D simulations in a weak stratification setup where
$\chi = 1.1$ and $m = 1.3$ in order to be close to the adiabatic
gradient. The initial perturbation is close to the fundamental
velocity mode. The spatial resolution is set to $128^2$, and we impose
the temperature flux on the bottom boundary. We then obtain stationary
symmetric convective rolls. We study the effect of the low-Mach
correction on the onset of the Rayleigh-B\'{e}nard instability by
varying the initial Rayleigh number. Figure~\ref{fig:growth_rates}
shows the evolution of the mean absolute velocity. The linear phase,
in logarithmic scale, corresponds to the exponential growth of
modes. We can see that without the low-Mach correction we have an
effective critical Rayleigh number between 10 and 15. Whereas with the
low-Mach correction we recover an effective critical Rayleigh number
close to the theoretical critical one.\par
\begin{figure}
  \centering
  \includegraphics{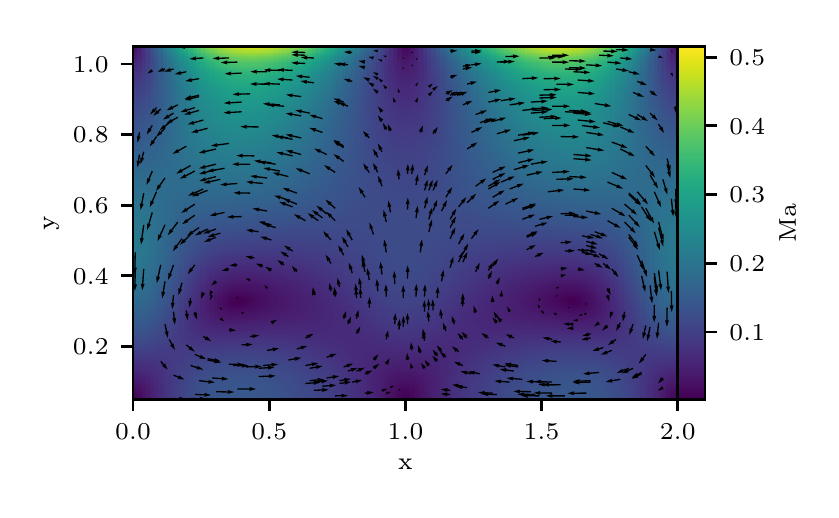}
  \caption{\label{fig:strong_stratification} Rayleigh-B\'{e}nard
    instability simulations. Snapshot of the local Mach number field
    and the velocity field. We see that in the strong stratification
    case, there is a large range of Mach, near zero at the center of
    rolls up to half at the upper boundary.}
\end{figure}

{}
If we now turn to a stronger stratification, the convective rolls
pattern change. We increase the density ratio to $\chi =
21$. Figure~\ref{fig:strong_stratification} shows a snapshot of the
local Mach number field with the velocity field, low-Mach correction
enabled. As observed in~\cite{hurlburt_two-dimensional_1984} we see a
downward shift of the center of mass of convective rolls compared to
the weak stratification case. By conservation of mass, the upper part
of the convective roll has to be larger. The strong stratification
case also exhibits higher Mach flows, around $\mathrm{Ma} \approx 0.5$
at the top of the box due to the low density. The all-regime well
balanced scheme is indeed able to capture properly convection in
highly stratified and high Mach flows.
\subsubsection{3D case}
\begin{figure}[!h]
  \centering
  \includegraphics[width=\linewidth]{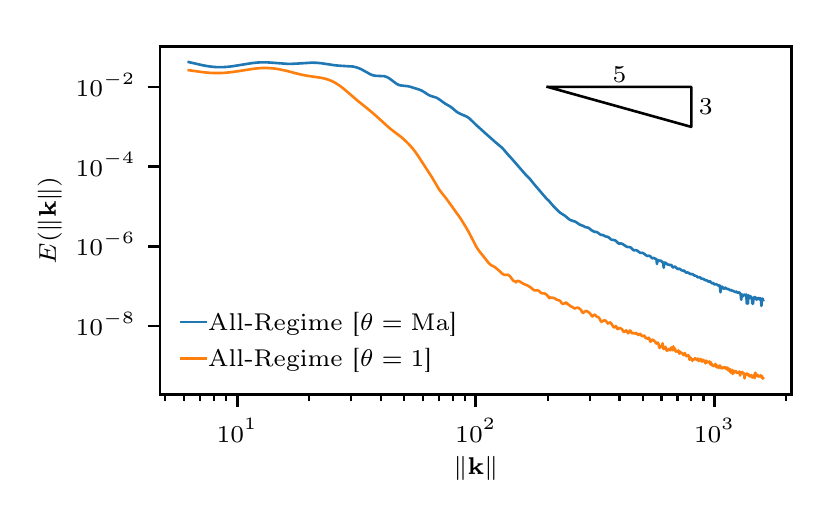}
  \caption{\label{fig:spectrum} Rayleigh-B\'{e}nard instability
    simulations in 3D. Figure shows the kinetic energy spectrum of the
    horizontal middle plane. The blue line corresponds to the scheme
    with low-Mach correction and the orange one without the low-Mach
    correction. \refereecorrection{We see more kinetic energy at all
      scales in the case of the low-Mach correction.}}
\end{figure}
\begin{figure*}
  \centering
  \includegraphics[width=\linewidth]{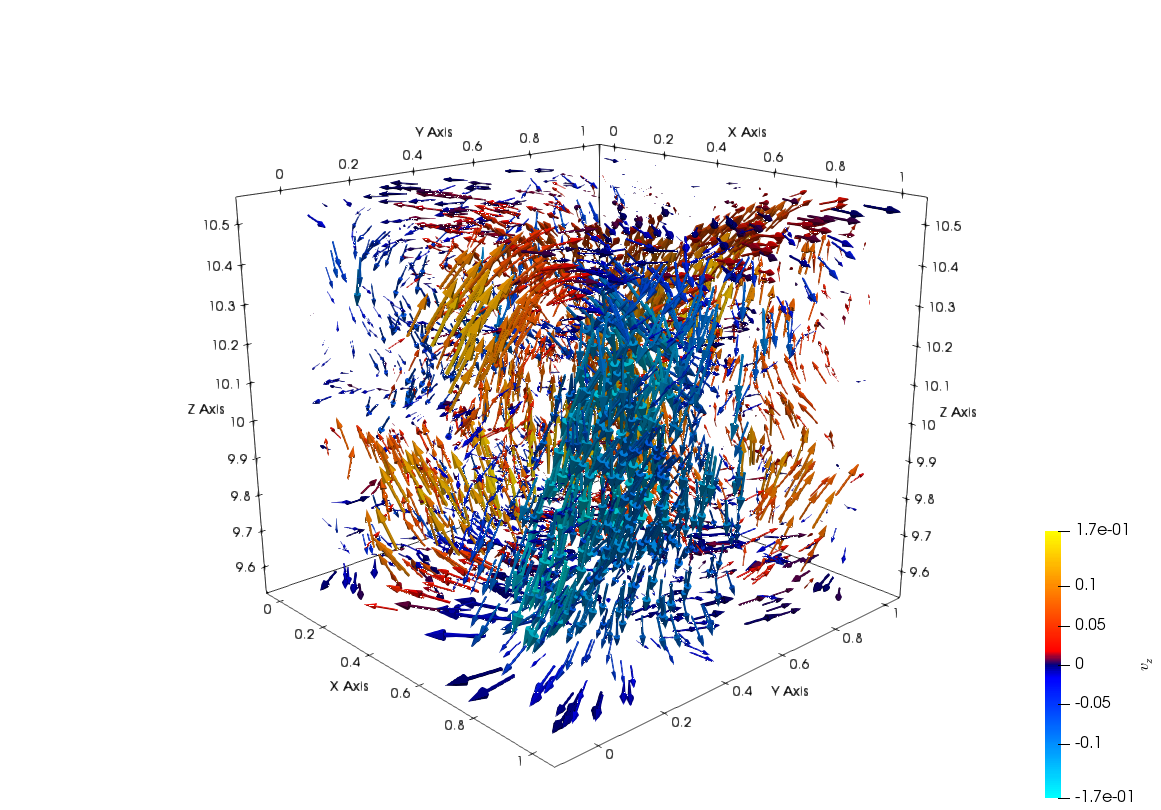}
  \caption{\label{fig:turbulence_snapshot} Rayleigh-B\'{e}nard
    instability simulations in 3D. Figure shows the velocity field in
    the box. The length of an arrow is scaled using the magnitude of
    the local velocity. The colorbar represents the vertical component
    of the velocity showing the direction of the flow.}
\end{figure*}

We now turn to 3D simulations in a weak stratification situation. In
this setup we want to look at the effect of the low-Mach correction on
the kinetic energy spectrum in a more turbulent situation. So we
change the polytropic index to $m = 0.1$ and increase the initial
Rayleigh number to $\mathrm{Ra} \approx 650000$. We also change the
boundary condition to a fixed temperature for both boundaries in order
to continuously force a large Rayleigh number in the simulation. By
using different upscaling, from $128^3$ to $512^3$ we reach a
stationary state~\footnote{The simulation outputs are available at
  \url{http://opendata.erc-atmo.eu}}. Figure~\ref{fig:turbulence_snapshot}
shows a snapshot of the velocity in the box. We see large and
structured vertical flows whereas in horizontal plans the flow is more
turbulent. In order to study the different scales and the energy in
this turbulent state we compute power spectrum of the kinetic energy
of the horizontal middle plane. Figure~\ref{fig:spectrum} shows the
results, the orange curve corresponds to the simulation performed with
the low-Mach correction and the blue one without it. We see a net
difference in the overall kinetic energy due to a lower dissipation
into the internal energy. \refereecorrection{We notice that we recover
  higher kinetic energies at all scales showing that the low Mach
  correction is important to properly capture the power spectrum of
  turbulent convection.}
\section*{Conclusion}
We have presented a new numerical code that is able to perform
simulations of convection without any approximation of Boussinesq nor
anelastic type. To do so we have adapted an all-Mach number scheme
into a well-balanced scheme for gravity. We have been able to show
that it preserves arbitrary discrete equilibrium states up to the
machine precision. Moreover the low-Mach correction in the numerical
flux allows to be more precised in the low-Mach regime. This new
scheme is well suited to properly study highly stratified and high
Mach convective flows. The low Mach correction is important to
properly capture convection modes in the laminar low Mach regime and
the kinetic energy power spectrum in the turbulent regime. This code
has been parallelized using a hybrid approach MPI+Kokkos in order to
be well prepared for running on forthcoming exascale machines.\par{}
Further work will consist in using the implicit-explicit approach to
reach very low Mach number simulations,
see~\cite{chalons_all-regime_2016}, and still keeping the
well-balanced property for the gravity source term. Indeed by solving
the acoustic part implicitly we avoid the restrictive CFL condition
due to the fast acoustic waves. With both the explicit-explicit and
implicit-explicit approach, this numerical scheme will be able to
efficiently study convection problems in all regimes, low Mach and
high Mach on the largest next generation massively parallel
architectures.
\section*{Acknowledgement}
\refereecorrection{P. Tremblin} acknowledges supports by the European
Research Council under Grant Agreement ATMO 757858. This work was
granted access to the HPC resources of TGCC under the allocation
A0040410097 attributed by GENCI (Grand Equipement National de Calcul
Intensif). \refereecorrection{The authors acknowledge IDRIS (Institut
  du D\'eveloppement et des Ressources en Informatique Scientifique)
  center to allow access to the Ouessant supercomputer.} The authors
would also like to thank Martial Mancip (CEA Saclay, Maison De La
Simulation) for helping in remote visualization, Maxime Stauffert for
his insight in the development of the well-balanced scheme and
\refereecorrection{G. Grasseau (LLR, Polytechnique, IN2P3) to allow
  access to a \uppercase{Nvidia} V100 GPU}.
\appendix
\section{Eigenstructure of the acoustic system}\label{appendix:eigenstructure}
For the sake of simplicity, the eigenstructure analysis of the
acoustic system~\eqref{eq:acoustic_subsystem} is made in the
one-dimensional case. We use the following change of variables, valid
for smooth flows
\begin{equation*}
  (\rho, \rho u, \rho \mathcal{E}, \rho \Psi, \Phi) \to (\rho, u, s, \Psi, \Phi),
\end{equation*}
where $s$ is the specific entropy. By
using equation of mass, one obtains
\begin{equation*}
  \begin{aligned}
    \partial_t \rho + \rho \partial_x u &= 0,\\
    \partial_t u + \frac{1}{\rho} \partial_x p^{\text{EOS}} + \partial_x \Phi &= 0,\\
    \partial_t e - \frac{p}{\rho} \partial_x u &= 0,\\
    \partial_t \Psi - u \partial_x \Phi &= 0,\\
    \partial_t \Phi &= 0.
  \end{aligned}
\end{equation*}
By using the second law of Thermodynamics and the equation on the
specific internal energy, one can show that
$\partial_t s = 0$~\citep[see][]{godlewski_numerical_1996}. Thus the
acoustic system~\eqref{eq:acoustic_subsystem} writes equivalently
\begin{equation}\label{eq:acoustic_system_primitive_variables}
  \begin{aligned}
    \partial_t \rho + \rho \partial_x u &= 0,\\
    \partial_t u + \frac{1}{\rho} \partial_x p^{\text{EOS}} + \partial_x \Phi &= 0,\\
    \partial_t s &= 0,\\
    \partial_t \Psi - u \partial_x \Phi &= 0,\\
    \partial_t \Phi &= 0.
  \end{aligned}
\end{equation}
The Jacobian matrix associated to the quasi-linear
system~\ref{eq:acoustic_system_primitive_variables} involves five
eigenvalues: $-c < 0 < c$ where 0 is degenerated three times and $c$
satifies
$c^2 = \partial_{\rho} p^{\text{EOS}} \left( \rho, s \right)$. It is
then hyperbolic. The four eigenvectors are given by
\begin{equation*}
  \vect{r}^0_0 = {\left(\partial_s p, 0, - c^2, 0, 0\right)}^{\mathrm{T}}, \quad \vect{r}^1_0 = {\left(0, 0, 0, 1, 0\right)}^{\mathrm{T}}, \quad \vect{r}_{\pm c} = {\left(\rho, \pm c, 0, 0, 0\right)}^{\mathrm{T}}.
\end{equation*}
Clearly the field associated to the stationary wave is linearly
degenerated. The fields associated to $\pm c$ are genuinely non-linear
under the condition that the following quantity does not vanish
\begin{equation*}
  \pm \nabla{c \left( \rho, s \right)} \cdot \vect{r}_{\pm c} = \pm \rho \partial_{\rho} c = \pm \frac{\rho}{2 c} \partial^2_{\rho \rho} p^{\text{EOS}}.
\end{equation*}
\section{Non-conservative energy
  scheme}\label{appendix:non-conservative}
To obtain the non-conservative scheme, we do not need anymore the
relaxation on the gravitational potential. This scheme is then
obtained through the following splitting, for the acoustic subsystem
\begin{equation*}
  \begin{alignedat}{5}
    &\partial_t \rho &&+ & \rho & &&\div{\vect{u}} &&= 0,\\
    &\partial_t \left( \rho \vect{u} \right) &&+ &\rho \vect{u}& &&\div{\vect{u}} + \grad{p} &&= - \rho \grad{\Phi},\\
    &\partial_t \left( \rho E \right) &&+ &\rho E& &&\div{\vect{u}} + \div{\left(p\vect{u}\right)}&&= - \rho \vect{u} \cdot \grad{\Phi},\\
  \end{alignedat}
\end{equation*}
followed by the transport subsystem
\begin{equation*}
  \begin{alignedat}{3}
    &\partial_t \rho &&+ \vect{u} \cdot \grad{\rho} &&= 0,\\
    &\partial_t \left( \rho \vect{u} \right) &&+ \vect{u} \cdot \grad{\left(\rho \vect{u}\right)} &&= \vect{0},\\
    &\partial_t \left( \rho E \right) &&+ \vect{u} \cdot
    \grad{\left(\rho E\right)} &&= 0,
  \end{alignedat}
\end{equation*}
then we use the same techniques for the acoustic system as
in~\ref{subsub:acoustic_system}, in other words the use of the mass
variable and the Lagrangian variables. The acoustic system in these
variables writes
\begin{alignat*}{3}
  &\partial_t \tau &&- \partial_m u &&= 0,\\
  &\partial_t u &&+ \partial_m p &&= - \frac{1}{\tau} \partial_m \Phi,\\
  &\partial_t v && &&= 0,\\
  &\partial_t E &&+ \partial_m (p u) &&= - \frac{u}{\tau} \partial_m
  \Phi,
\end{alignat*}
\begin{equation*}
    E = e + \frac{1}{2} (u^2 + v^2).
\end{equation*}%
Using a pressure relaxation, an approximate Riemann solver with source
term, see~\cite{gallice_solveurs_2002}, and the same upwind scheme for
the transport system as in~\ref{subsub:transport_system} we obtain the
following non-conservative counterpart scheme
\begin{alignat*}{2}
  &\rho^{n+1}_i &&= \rho^n_i - \frac{\Delta t}{\Delta x} {\left[ \widetilde{\rho} u^* \right]}_i,\\
  &{\left(\rho u\right)}^{n+1}_i &&= {\left(\rho u\right)}^n_i - \frac{\Delta t}{\Delta x} {\left[  \widetilde{\left(\rho u\right)} u^* + \Pi^* \right]}_i + \frac{\Delta t}{\Delta x} S^n_i,\\
  &{\left(\rho v\right)}^{n+1}_i &&= {\left(\rho v\right)}^n_i - \frac{\Delta t}{\Delta x} {\left[  \widetilde{\left(\rho v\right)} u^*\right]}_i,\\
  &{\left(\rho E\right)}^{n+1}_i &&= {\left(\rho E\right)}^n_i - \frac{\Delta t}{\Delta x} {\left[ \left(\widetilde{\left(\rho E\right)} + \Pi^*\right) u^* \right]}_i + \frac{\Delta t}{\Delta x} {\left( u S \right)}^n_i,
\end{alignat*}
where
\begin{align*}
  u^*_{i+1/2} &= \frac{1}{2} (u^n_{i+1} + u^n_i) - \frac{1}{2 a} \left( \Pi^n_{i+1} - \Pi^n_i - S^n_{i+1/2}\right),\\
  \Pi^*_{i+1/2} &= \frac{1}{2} \left(\Pi^n_{i+1} + \Pi^n_i\right) - \frac{a}{2} \left(u^n_{i+1} - u^n_i\right),\\
  a^n_{i+1/2} &\geq \max{\left(\rho^n_i c^n_i, \rho^n_{i+1} c^n_{i+1}\right)},\\
  S^n_i &= \frac{1}{2} \left( S^n_{i+1/2} + S^n_{i-1/2} \right),\\
  {(uS)}^n_i &= \frac{1}{2} (u^*_{i+1/2} S^n_{i+1/2} + u^*_{i-1/2} S^n_{i-1/2}),\\
  S^n_{i+1/2} &= - \frac{1}{2} \left( \frac{1}{\tau^n_i} +
                \frac{1}{\tau_{i+1}} \right) \left( \Phi_{i+1} - \Phi_i \right).
\end{align*}
This scheme is not conservative for the whole energy but is closer to
the scheme proposed for the shallow water equations
in~\cite{chalons_large_2017}, for which the authors have obtained a
discrete entropy inequality. It seems therefore possible to obtain a
similar inequality for this non-conservative scheme, but this
demonstration is beyond the scope of this paper.
\bibliographystyle{aasjournal}
\bibliography{main}
\end{document}